# EVALUATION OF ENERGY CONSUMPTION OF REACTIVE AND PROACTIVE ROUTING PROTOCOLS IN MANET

Mohamad T. Sultan and Salim M. Zaki

Department of Computer Science, Cihan University, Erbil, Iraq

*ABSTRACT*

*Mobile Ad hoc Network (MANET) is a distributed, infrastructure-less and decentralized network. A routing protocol in MANET is used to find routes between mobile nodes to facilitate communication within the network. Numerous routing protocols have been proposed for MANET. Those routing protocols are designed to adaptively accommodate for dynamic unpredictable changes in network's topology. The mobile nodes in MANET are often powered by limited batteries and network lifetime relies heavily on the energy consumption of nodes. In consequence, the lack of a mobile node can lead to network partitioning. In this paper we analyse, evaluate and measure the energy efficiency of three prominent MANET routing protocols namely DSR, AODV and OLSR in addition to modified protocols. These routing protocols follow the reactive and the proactive routing schemes. A discussion and comparison highlighting their particular merits and drawbacks are also presented. Evaluation study and simulations are performed using NS-2 and its accompanying tools for analysis and investigation of results.*

## 1. INTRODUCTION

The advent of wireless mobile ad-hoc networks (MANETs) has offered an efficient and most importantly cost effective technique to make use of the availability of mobile hosts when no fixed infrastructure is provided. In MANET, the mobile nodes can easily communicate with each other while they are freely moving around in different directions. An ad-hoc network relies entirely on nodes cooperation for forwarding information from data sources to intended destination nodes. Some examples of mobile nodes in an ad-hoc network are laptop computers, smart phones and personal digital assistants that interact directly with each other [1][16].There are many advantages of such an ad-hoc network which include fast deployment, robustness, efficiency, and inherent support for mobility.

The mobile nodes in MANET can be arbitrarily positioned and are free to travel randomly at any particular time, thus allowing network topology and connections between mobile nodes to change rapidly. This makes routing in MANET a challenge. Routing protocols helps MANET to perform its function of routing of the data packets from the source to the intended destination in the network. Because of the occurrence of mobility, the routing information will have to be changed to reflect changes in link connectivity [1][2].

Researchers have become more interested on how to secure MANETs. They have suggested several methods to prevent or reduce the risk of attacks on the mobile ad-hoc networks by using access control key management and trust models mechanisms [1][2].Energy consumption is a very critical issue in MANET since mobile nodes are often powered by limited battery resources. Thus, network lifetime relies heavily on the energy consumption of nodes. Saving energy is, therefore, critical in order to prolong the lifetime of the network. Figure 1 illustrates the architecture of the Mobile Ad-hoc Network (MANET).

The paper is organized as follows. Section 2 reviews the related work. Section 3 briefly describes the studied routing protocols. Section 4 gives the details of simulation environment and energy model. It also describes the implementation of the routing protocols and the simulation setup used in this research. The Simulation results are shown in section 5. Finally section 6 describes our conclusion and future scope.

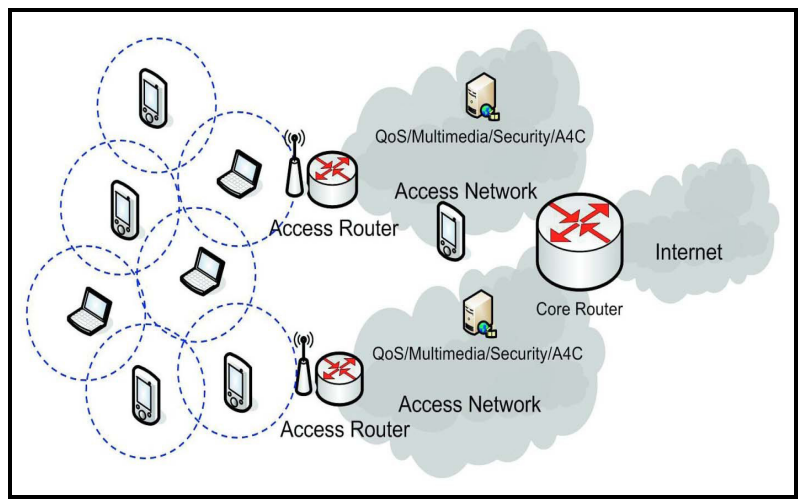

Figure 1. MANET architecture

## 2. RELATED WORK

This section summarizes some of the work related to MANET routing protocols in terms of energy consumption done by several researchers. The authors in [3] have proposed a loop-free energy conserving scheme which tries to decrease routing and storage overhead to provide optimization of resources use in large scale networks. The researchers in [4] have found reactive protocols such as DSR; AODV behaved more efficient than DSDV and showed superior performance than TORA. In [5] the authors have investigated AODV based algorithm with less energy consumption during route founding by establishing routes that are lower congested than the others. The traditional routing algorithms lack energy awareness of the nodes in the network [17].

The authors in [20] have compared the energy consumption of various protocols under CBR traffic. The authors in [21] have compared two reactive protocols under ON/OFF source traffic. They have selected packet delivery ratio, normalized routing overhead and average delay as the performance parameters. Jaun Carlos Cano et. al. [22] have developed number of such protocols and analyzed them under Constant Bit Rate (CBR) traffic. D. Nitnawale et. al. [23] have presented a paper on comparison of various protocols under Pareto traffic. An analysis of these studies shows that their shared goal is to enhance the energy consumption of routing protocols. However, the parameters taken into consideration by each of them are different. In the current paper, we have evaluates and analyzed the energy consumption of three prominent MANET routing protocols (AODV, OLSR and DSR) under constant bit rate CBR traffic in regards to different number of mobile nodes. Total energy consumed by each node throughout transmission and reception operation has been evaluated as the function of number of nodes in the network.

## 3. MANET ROUTING PROTOCOLS

Several routing protocols for ad-hoc networks have been proposed. The two main basic roles of MANET routing protocols are the selection of path and delivering data packets accurately to the right target. MANET routing protocols can be classified into three different categories which are table-driven routing protocols (proactive), on-demand routing protocols (reactive) and hybrid routing protocols [6][15]. In the following sub-sections, the proactive and the reactive routing approaches will be discussed.

### 3.1 Table-Driven Routing Protocols (Proactive)

The table-driven routing protocols always try to keep consistent up to date routing data for every node in the network. These protocols attempt to maintain accurate routing information of the complete network at all times. Every node in the network keeps the routing information by maintaining one or more routing tables. The nodes usually try to revitalize the information about the target nodes by updating the routing tables. The routing protocol adapts to the sudden changes in topology by broadcasting network updates whenever changes occur [6]. In the following section an explanation of OLSR routing protocol will be presented.

### 3.1.1 Optimized Link State Routing (OLSR)

In OLSR all nodes have routing table for keeping the routing information to every other node in the network to provide a route to the destination immediately when desired [7]. In OLSR routing protocol, one of the most essential key concepts used is the use of multipoint relays (MPRs). The main purpose of the MPRs is to forward the broadcast messages in the network. The traditional link state protocol broadcast mechanism is not used in OLSR. This is because in OLSR only partial link state information is distributed where the content of the broadcast packets is only about MPRs rather than all the detailed link state information [8].

The mobile nodes select the MPRs amidst their surrounding neighbors and then rebroadcast only those messages that are received from nodes who selected it as an MPR. OLSR mainly uses two kinds of control messages. The first one is the periodic HELLO messages while the second type is the Topology Control (TC) messages. The HELLO messages are used for discovering the information about the link status or in other words it carries out the task of neighbor detecting. The second type, which is the Topology Control (TC) messages are used for the purpose of information declaration about the multipoint relay. OLSR protocol is mainly appropriate for large and dense networks for the reason that the technique of multipoint relays performs well in this environment [7][8]. Figure 2 below shows MPR selection in OLSR routing protocol.

### 3.1.2 ENERGY EFFICIENT OLSR ROUTING PROTOCOL (EE-OLSR)

uses multipoint relays (MPRs) to elect nodes for forwarding messages the protocol uses energy aware mechanism called willingness which is a variable representing the availability of a node to act as a MPR for surrounding nodes. The idea behind this method is to reduce the messages overhead that leads to reduction in energy consumption. The main goal of EE-OLSR is to prolong the lifetime of the ad hoc network by extending lifetime of nodes' batteries.  As shown in simulation results, EE-OLSR protocol overtakes OLSR protocol in average nodes lifetime and preserving the normalized control overhead [18].

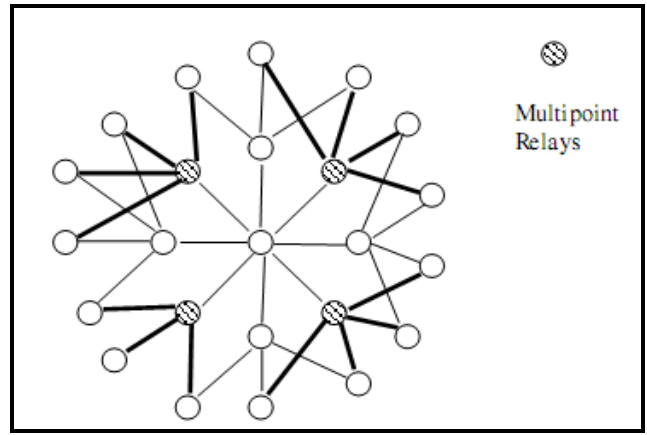

Figure 2. Multipoint relays in OLSR routing protocol

**3.2 On-Demand Routing Protocols (Reactive)**

On-demand protocols generate routes only when there is a request to send data. This approach is totally different than that of table-driven routing. Routing information about the mobile nodes is not maintained by protocols instead the route request is initiated only when needed. When the source nodes need to start a route to the target, at this time the route is being initiated to serve the request of the source nodes [6][9].

**3.2.1 Ad-hoc On-Demand Distance Vector (AODV)**

AODV is developed based on the DSDV routing algorithm [6][10]. It is categorized as a pure reactive routing protocol scheme. When there is a route required in the network, AODV will execute route discovery process to find the route to the desired destination. As soon as a route is generated in the network, it is maintained as long as it is still needed using a route maintenance process. Each mobile node keeps the routes that have been discovered in its routing table. However a routing table entry expires if it is no longer being used or has exceeded the expiration period which has been pre-identified earlier. AODV has three essential messages which are the Route Request (RREQ), Route Reply (RREP), and Route Error (RERR) [6]. The RREQ message contains information such as the IP address of the source and destination nodes, current sequence number, broadcast ID and latest sequence number for the destination known to the source node. In response to the request message the mobile nodes send back RREP message to the source node. This reply message contains the information that the source node needs with the valid route. However, when there is a problem in the network another procedure is used which is the RERR message. This message holds a list of all of the unreachable destinations in the network. AODV is a reactive on-demand routing protocol, despite the fact that it still employs some characteristics of the table-driven routing approach. AODV is a good choice for routing in the case when the network is dynamically changing. This protocol combines the motivating features of DSR and DSDV routing in a way that it employs the idea of route discovery and route maintenance like DSR routing protocol and in the same time it makes use of the sequence numbers and transmitting of periodic hello messages from DSDV. The use of destination sequence numbers guarantees loop-free routes and help to detect available fresh routes which allow the source nodes to discover new routes every time[11].

### 3.2.2 Energy aware routing (EAR)

This protocol is based on AODV it aims to make the network energy aware. Where energy efficient design of the protocol is controlled through changing the transmission range of the nodes. Adjustable transmission range affects the battery power for each packet at each node, accordingly affects energy consumption of the ad hoc network. The higher transmission range the less number of hops to destination node, the less transmission range the high number of forwarders to destination node. The high transmission affects the power negatively and leads to high power consumption. EAR utilizes Route Discovery phase when node communication with neighbours. Once the route is determined, every node controls the transmission range as per the distance between source and destination node, therefore the best energy is utilized for packet transmission. The simulation results show EAR outperforms AODV in terms of network time and energy consumption [17].

### 3.2.3 Dynamic Source Routing (DSR)

DSR is an independent routing protocol which can be described as a completely self-configuring self-organizing routing protocol [12]. To improve the discovery process in the network, routes caches are kept at the mobile nodes and those caches are frequently updated. DSR involves no periodic data packets within the network of any kind at any level. For instance, there is no periodic routing advertisement, neighbor detection or link status sensing packets to be used in DSR, and it does not depend on any underlying protocol for those tasks. DSR has two core mechanisms which are route discovery and route maintenance [13]. Once any mobile node within the network desires to send a data message to a particular destination, it initially broadcasts a route request (RREQ) packet. The neighbor and surrounding nodes which lay in the range of transmission of the source node receive that RREQ packet and add their own address to it and they rebroadcast it again in the network. If the discovery procedure is successful, the source initiator obtains a response data packet that shows the series of nodes over which the destination could be reached. The route request packet therefore has a record field accumulating a list of nodes visited for the duration of propagation of the query in the network. DSR routing protocol has many benefits. For instance, it does not employ periodic route advertisement which leads to saving in network bandwidth as well as reduction in power consumption. DSR has a faster route recovery than many other reactive protocols as well. However, the limitation of this protocol is that the benefit of caching routes for large networks and higher mobility may become not that useful [12][14].

### 3.2.4 Multipath and energy-aware on demand source routing (MEA-DSR)

This protocol is based on DSR. MEA-DSR, minimizes the number of discovered paths that a destination node provides to a source node to two paths only. The criteria of choosing primary route in MEA-DSR is conditioned by two features; first, the remaining energy of nodes belonging to the path; second, the total transmission power essential to transmit data on the path. This feature is corresponding to that of number of hops in the route if we assume that nodes transmit with full power. Multiple modifications done on messages of DSR to make MEA-DSR energy aware and select reliable paths. The simulation results show better performance of MEA-DSR compared to DSR in consumed energy [19].

## 4. SIMULATION ENVIRONMENT

This section provides the details of the simulation environment used in this paper. The entire simulation work is conducted and implemented on a Linux (Ubuntu distribution) operating system. The simulation is done with the help of NS-2 simulator. NS-2 deals efficiently with network's core components and it provides the complete vision of the network construction this includes routing protocols, transport layer protocols, interface queues, as well as link layer components. The simulation environment consists of four different numbers of nodes which are 10, 20, 30 and 50 mobile nodes. Nodes are being generated randomly at random position and Constant Bit Rate (CBR) traffic generators will be used as sources to run the simulation. For this research study the selected mobility model is the Random Waypoint Mobility Model which is one of the most widely used mobility models among the research community. The selected parameters are varied using setdest command in NS-2.

The simulation parameters considered for the performance evaluation of MANET routing protocols are shown in Table 1.

Table 1. Simulation Parameters

| Simulation parameters | |
|---|---|
| **Parameters** | **values** |
| Platform | Linux (Ubuntu) 10.04 |
| Simulation Tool | Network Simulator 2 (NS-2) |
| Routing Protocols | AODV, DSR and OLSR |
| Pause Time | 10 sec. |
| Experiment Duration | 130 sec |
| Number of Nodes | 10, 20, 30, 50 |
| Traffic Model | CBR (Constant Bit Rate) |
| Packet Size | 512 bytes |
| Area | 500m X 500m |
| Maximum Speed | 20 m/s |
| Mobility Model | Random Waypoint |
| MAC Layer Protocol | IEEE 802.11b |
| Antenna Type | Antenna/OmniAntenna |

### 4.1 Energy Evaluation Model

We have used similar energy model as specified by Marzoni and Cano [4]. Energy is converted in joules by multiplying power with time. The total energy consumed by each node is calculated as sum of transmitted and received energy for all control packets. The following equations are used to calculate energy required in joules to transmit and receive the packets of given size in joules.

Transmitted Energy:

Tx Energy = (Tx Power X Packet Size) / $2\times10^6$   (1)

Receiving Energy:

Rx Energy = (Rx Power X Packet Size) / $2\times10^6$   (2)

The table below show the energy model and energy parameters used in this study.

Table 2. Energy Model Parameters

| Parameters for Energy Model | |
|---|---|
| **Parameters** | **values** |
| Initial Energy | 100 Joule |
| Radio Frequency | 281.8mW |
| Receiving Power | 1.1w |
| Transmission Power | 1.65w |
| Transition Power | 0.6w |
| Sleep Power | 0.001w |
| Transition Time | 0.005s |

## 5. SIMULATION RESULTS AND ANALYSIS

The simulation results of each network scenario that presents the performance of the routing protocols with respect to energy consumption are explained in this section. The node density of the simulation network was varied to determine the performance impact of the three routing protocols.

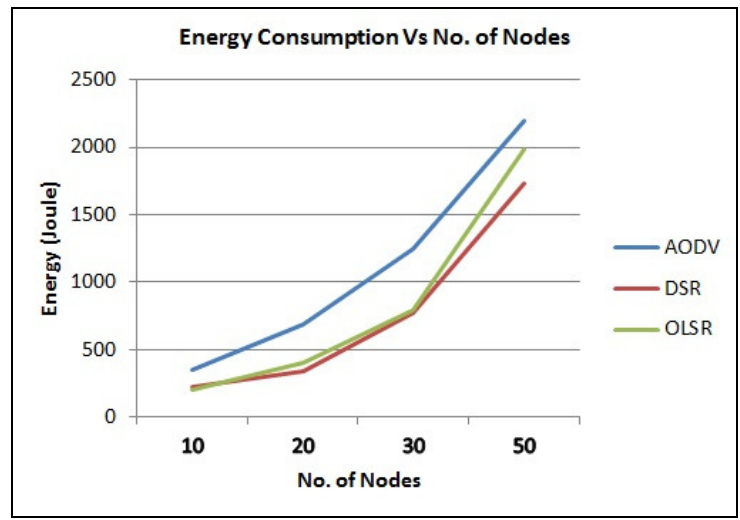

Figure. 3 Energy consumption Versus No. of Nodes

Figure 3 shows the total energy consumed (Joules) by all the nodes involved in transmitting and receiving the control packets by varying the number of mobile nodes in the network. The node density is increased from 10 to 50 nodes within the same network size as each simulation run is performed.

All the routing protocols show an increase in the consumed energy as the number of the mobile nodes increase and the network size become bigger. This is because the mobile nodes in the network have to process all the routing packets. As such, the total power consumption of the network will increase. According to Figure 3, the consumed energy of DSR and AODV and OLSR form 10 nodes to 20 nodes are quite similar; but a considerable difference of energy consumption can be noticed as the network size become larger until it reaches to 50 mobile nodes. The performance of the on-demand routing protocol AODV in terms of the energy consumption is quite poor compared to the other on-demand protocol DSR, and this could be mainly due to the increase in the maintenance process as the number of nodes increase. Although DSR and AODV have the same on-demand behavior but they are still have a bit different routing mechanisms. That's why AODV has superior energy consumption as compared to DSR where maintenance process of AODV can be the main reason of this increase.

The proactive routing protocol OLSR has an average performance compared to the other protocols. From the figure above it can be seen that the energy consumption of both DSR and OLSR routing protocols increases in quite the same pattern with increasing number of nodes below 30; but the gap in energy consumption only start to occur half way through the simulation and start to become higher after 30 mobile nodes. The short comings of this aspect of OLSR are caused by the large number of overheads generated between the nodes within a group. However, OLSR did not perform too badly and has consistently better results than AODV. At higher density nodes, DSR routing protocol performs efficiently in consuming less overall system energy and this is because of DSR as an on-demand protocol doesn't have to maintain route to the target if there is no data to be send. That's why DSR seems to have better performance compared to its counterparts.

## 6. CONCLUSION

In this paper, we addressed the issues of energy efficient routing in MANET. Mobile nodes in MANET rely on batteries, consequently efficient utilization of battery energy becomes significant and it affects the increase the lifetime of the mobile network. This paper is mainly focused on several routing algorithms proposed for mobile ad-hoc networks (MANETs). The aim of this research work was to evaluate the performance of three prominent MANET routing schemes which are AODV, DSR and OLSR with regard to energy consumption. The purpose of the work is to comprehend the impact of having more nodes within a fixed map of operation on the network's energy consumed. In general, the on-demand AODV protocol consumed more energy within the network at the beginning of the simulation and it lasted until the end of the simulation. As mentioned earlier AODV routing protocol has shown higher degree of consumed energy than the DSR and OLSR routing in higher density network operation. While DSR uses source routing with longer header, AODV uses hop-by-hop scheme which may not help in consuming lower energy values.

It observed generally that increasing number of nodes results in increasing energy consumption in all routing protocols due to routing control packets. It can be noticed of DSR as being more energy efficient than AODV and OLSR in this research and it emerges as a good candidate in conserving the system energy. This work may generally be concluded by addressing the issue of routing in MANETs which is each routing protocol has certain benefits and weaknesses, and is well suited for certain conditions. Therefore choosing an appropriate routing protocol for MANET would produce the highest routing performance in the network.